\def\draftversion{true}
\def\draftversion{false}
\RequirePackage{ifthen}
\ifthenelse{\equal{\draftversion}{true}}{
  \documentclass[aps,prb,10pt,galley,amsmath,amssymb,
                 superscriptaddress,citeautoscript]{revtex4}
}{
  \documentclass[aps,prb,10pt,twocolumn,amsmath,amssymb,longbibliography,
                 superscriptaddress,citeautoscript]{revtex4-1}
}

\usepackage{graphicx}
\usepackage[usenames,dvipsnames]{color} 
\usepackage{bm} 
\usepackage{soul} 

\usepackage{xspace} 

\ifthenelse{\equal{\draftversion}{true}}{
  \marginparwidth 2.7in
  \marginparsep 0.5in
  \newcounter{comm} 
  \def\commnext{\stepcounter{comm}}
  \def\commtext{{\bf\color{blue}[\arabic{comm}]}}
  \def\commmar{{\bf\color{blue}[\arabic{comm}]}}
  \def\dvm#1{\commnext\marginpar{\small DV\commmar: #1}\commtext}
  \def\sbm#1{\commnext\marginpar{\small SB\commmar: #1}\commtext}
  \def\mlab#1{\marginpar{\small\bf #1}}
  
}{
  \def\dvm#1{}
  \def\sbm#1{}
  \def\mlab#1{}
  
}
\newcommand{\beq}{\begin{equation}}
\newcommand{\eeq}{\end{equation}}
\newcommand{\bea}{\begin{eqnarray}}
\newcommand{\eea}{\end{eqnarray}}

\newcommand{\eqlab}[1]{\label{eq:#1}}
\newcommand{\eq}[1]{Eq.~(\ref{eq:#1})}

\newcommand{\figlab}[1]{\label{fig:#1}}
\newcommand{\fref}[1]{Fig.~\ref{fig:#1}}
\newcommand{\frefs}[1]{Figs.~\ref{fig:#1}}

\newcommand{\Frefs}[1]{Figures~\ref{fig:#1}}
\newcommand{\ket}[1]{\vert#1\rangle}
\newcommand{\bra}[1]{\langle#1\vert}

\def\z2{$\mathbb{Z}_2$}

\def\k{{\bf k}}

\def\kdotp{$k\cdot p$\xspace}

\begin{document}



\title{First principles theory of Dirac semimetal Cd$_3$As$_2$
under Zeeman magnetic field}

\author{Santu Baidya}
\affiliation{
Department of Physics \& Astronomy, Rutgers University,
Piscataway, New Jersey 08854, USA}

\author{David Vanderbilt}
\email{dhv@physics.rutgers.edu}
\affiliation{
Department of Physics \& Astronomy, Rutgers University,
Piscataway, New Jersey 08854, USA}

\date{\today}
\begin{abstract}
Time-reversal broken Weyl semimetals have attracted much attention
recently, but certain aspects of their behavior, including
the evolution of their Fermi surface topology and anomalous Hall conductivity
with Fermi-level position, have remained underexplored.
A promising route to obtain such materials may be to start with
a nonmagnetic Dirac semimetal and break time-reversal symmetry
via magnetic doping or magnetic proximity.  Here we explore this scenario
in the case of the Dirac semimetal Cd$_{3}$As$_{2}$, based on
first-principles density-functional calculations and subsequent
low-energy modeling of Cd$_{3}$As$_{2}$ in the presence of a Zeeman
field applied along the symmetry axis.  We clarify how each
four$-$fold degenerate Dirac node splits into four
Weyl nodes, two with chirality $\pm 1$
and two higher-order nodes with chirality $\pm 2$. Using a
minimal \kdotp model Hamiltonian whose parameters are fit to
the first-principles calculations, we detail the
evolution of the Fermi surfaces and their Chern numbers
as the Fermi energy is scanned across the region of
the Weyl nodes at fixed Zeeman field. We also compute the
intrinsic anomalous Hall conductivity as a function of
Fermi-level position, finding a characteristic inverted-dome
structure. Cd$_{3}$As$_{2}$ is especially well suited to such a
study because of its high mobility, but the qualitative behavior
revealed here should be applicable to other Dirac semimetals
as well.
\end{abstract}

\maketitle


\section{Introduction}
%
The compound Cd$_{3}$As$_{2}$ has been widely studied in recent
years for its three-dimensional graphene-like characteristics.
\cite{Z.Phys.Chem.B28.427.1935,PRM.2.120302.2018,PRB.88.125427.2013,
Nature.Mater.13.677681.2014,Nature.Mater.13.851856.2014,PRL.113.027603.2014}
The existence of three-dimensional Dirac cones at the Fermi level in this
compound has attracted much attention in the field of topological
semimetals, as the only Dirac semimetals observed experimentally to date
are Cd$_{3}$As$_{2}$\cite{Inorg.Chem.53.4062} and
Na$_{3}$Bi\cite{PhysRevB.85.195320,Liu864}. 
Cd$_{3}$As$_{2}$ has many interesting properties in addition
to the existence of the Dirac cone, such as an
abnormally large $g$-factor of around
$20$\cite{doi:10.1002/pssb.2220920106}, which still demands microscopic
understanding. Most of the interest in the past few years has focused
on the Dirac crossing, which is protected by the crystalline
$C_{4v}$ symmetry.

Starting from a Dirac Hamiltonian, a Weyl semimetal phase\cite{Weyl1929}
can be reached by breaking either
inversion symmetry or time-reversal (TR)
symmetry\cite{PhysRevLett.107.127205}. TR symmetry can be
broken by doping with magnetic ions, by proximity effects near an interface
to a magnetic material, or by application of an external magnetic
field.

In the first two cases, the TR breaking is most naturally represented
in terms of an effective Zeeman field acting on the spins,
while the last also brings in orbital effects.
A previous study of the Dirac semimetal Na$_{3}$Bi and
its evolution into a Weyl semimetal phase under Zeeman
field has been discussed using tight-binding methods in
Ref.~[\onlinecite{PRB.98.075123.2018}], but this material may not
be an optimal choice for such a study in view of its chemical
instability.

Regarding orbital effects in Cd$_{3}$As$_{2}$, the quantum
Hall effect in a nanoplate of thickness $50-100$\,nm has
been experimentally reported and attributed to Weyl orbit
formation\cite{Zhang2019}.
Thus, the appearance and evolution of the Weyl nodes in Dirac
materials with TR-broken perturbations is a topic of pressing
interest.

In this paper we focus on the effects of a Zeeman field
on the nodal structure, Fermi-surface configuration, and
anomalous Hall conductivity
of Cd$_{3}$As$_{2}$.  Our work is motivated in part by a recent
proposal\cite{PhysRevLett.120.067003,sun-lee-li}
that a Dirac semimetal could be a platform for the
realization of unconventional superconductivity in a TR-broken
Weyl semimetal\cite{sun-lee-li} resulting from the presence of
magnetic dopants or proximity to a magnetic substrate or overlayer.

We start from realistic first-principles density functional
theory (DFT) calculations in which a Zeeman field is applied along
the symmetry axis, and then construct a linearized \kdotp
model to describe the low-energy physics near the Dirac point.
While elementary discussions often describe the TR symmetry
breaking as resulting in a splitting of the Dirac node into a
pair of Weyl nodes, we clarify that four Weyl nodes appear
instead.
For a given strength of Zeeman field, we give a detailed
description of the evolution of the Fermi surfaces of the
electron and hole pockets, and their nontrivial Chern numbers,
as the Fermi level $E_F$ is tuned over the range of energies
where the Weyl points occur.
We also compute and track the anomalous Hall conductivity,
paying special attention to its behavior as $E_F$ passes through
the Weyl node positions, predicting a characteristic signature that
we suggest as a target of future experimental observation.

\begin{figure}
\includegraphics[scale=0.36]{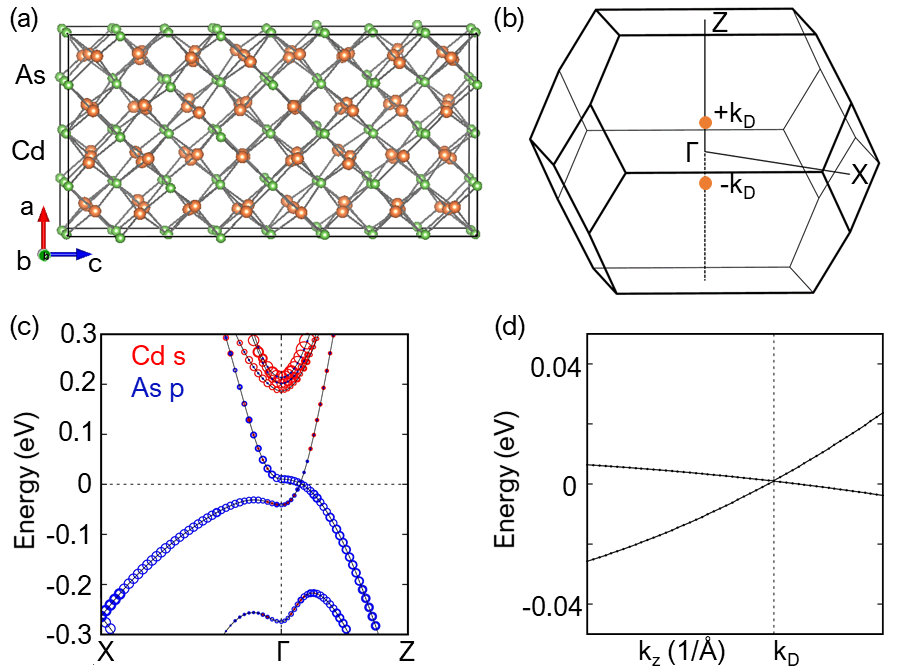}
\caption{
(a) Conventional unit cell of Cd$_{3}$As$_{2}$ with space group $I4_{1}/acd$. (b) Brillouin zone of the primitive unit cell with position of Dirac points. (c) Cd $s$ (red circles) and As $p$ (blue circles) character projected onto the energy manifold under nonmagnetic PBE+SOC approximation. (d) PBE+SOC band structure with linear Dirac crossing at the Fermi level}
\figlab{dft_band}
\end{figure} 

\section{Electronic structure of
C\lowercase{d}$_3$A\lowercase{s}$_2$ under Zeeman field}
%
\subsection{First-principles calculations}
%
Figure~\ref{fig:dft_band}(a) shows the crystal structure of the
Cd$_{3}$As$_{2}$, corresponding to a defective antifluorite
structure.
The compound has an intermediate-temperature phase
above 475$^{\circ}$C with space group
P4$_{2}$/nmc\cite{Z.Phys.Chem.B28.427.1935} having 40 atoms
in the primitive cell.
A high-temperature Fm$\bar 3$m antifluorite phase has
also been reported above 600$^{\circ}$C.
There has been confusion in the
literature about the low-temperature phase that occurs below
475$^{\circ}$C\cite{Yi2014,Inorg.Chem.53.4062,Steigmann:a06211}.
A previous work
identified the low-temperature phase as a noncentrosymmetric
I4$_{1}$cd
structure\cite{Yi2014,Steigmann:a06211}, and
subsequent theoretical tight-binding
calculations were carried out on this structure to propose a
possible TR-symmetric Weyl semimetal phase with broken inversion
symmetry.\cite{PRB.88.125427.2013}
However, recent experiments\cite{Inorg.Chem.53.4062} clarify
that this phase is indeed centrosymmetric, with space group
I4$_{1}$/acd containing 80 atoms per primitive cell.
We focus on the latter structure here.
The conventional 160-atom cell is shown in
\fref{dft_band}, and consists of eight layers of Cd atoms and eight layers
of As atoms stacked along the $c$ axis.
The structure can be regarded as an anti-fluorite structure with
$25\%$ vacancies on the Cd sites, with each As atom having six Cd
nearest neighbors.

DFT calculations are carried out in a full-potential linear augmented
plane wave (FP-LAPW) framework as implemented in the
Wien2k package\cite{wien2k}.  The
Perdew$-$Burke$-$Ernzerhof exchange-correlation functional
\cite{PhysRevLett.77.3865} is employed. Spin-orbit coupling (SOC)
is taken into account using the second-order variational approach
implemented in the Wien2k package.

The PBE band structure in \fref{dft_band}(c) shows the nonmagnetic band
structure near the Fermi level. All bands are doubly degenerate because
of the presence of inversion and TR symmetry.
Two of these degenerate bands cross each other at a
Dirac point (fourfold degeneracy) protected by $C_4$ symmetry
at $k_{z}=+k_{\rm D}$ on the
$\Gamma$-Z path along the $k_{z}$ axis.  The Cd
$s$ and As $p$ orbital projections shown in \fref{dft_band}(c) indicate
that the crossing occurs between an $s$-band with
positive slope and a $p$ band with negative slope. By symmetry,
there will be another Dirac point at $k_{z}=(0,0,-k_{\rm D})$ as shown
in Fig.~\ref{fig:dft_band}(b).
The zoomed view in Fig.~\ref{fig:dft_band}(d) shows the crossing point
along the positive $k_{z}$ axis.

To study the Weyl semimetal phase of Cd$_{3}$As$_{2}$, we calculate
the band structure under an effective Zeeman field
introduced to represent the effect of doping with magnetic
impurities at a mean-field level.  This is similar to the
spirit of previous works such as the study of
the quantum anomalous Hall effect in Cr-doped
topological insulator films of Bi$_{2}$Te$_{3}$\cite{yu-s10,Chang167}.
We anticipate collinear easy-axis ferromagnetic order, and thus apply
the field  along the $\hat{\bf z}$ symmetry axis.
While our theory treats the effective Zeeman
field $h$ as a free parameter,
we have chosen to present our results for
$h=100$\,T, corresponding to a splitting of 
11.6\,meV for a free electron, which could be a reasonable
spin-exchange field achievable by magnetic doping.
Orbital magnetic effects, such as those that give rise to
magnetoresistance oscillations, are not considered in our theory.

The effect of the Zeeman field on the band structure of the Dirac
semimetal is presented in Fig.~\ref{fig:dft_mag}.
For reference, \fref{dft_mag}(a) shows the band dispersion along the
$\Gamma$--$Z$ symmetry axis in the absence of the Zeeman field.
The Dirac crossing occurs at $k_z=k_{\rm D}=0.037$\,\AA$^{-1}$,
and by definition at zero energy.
Henceforth we
also reset the origin of $k_z$ to coincide with the Dirac
point location $k_{\rm D}$, so that subsequently $k_z$ is
always measured relative to $k_{\rm D}$.
The dispersions of the two crossing bands are roughly quadratic
relative to $\Gamma$, but show a linear crossing character when
attention is focused sufficiently close to $k_{\rm D}$.

Figure~\ref{fig:dft_mag}(b) shows the corresponding band
structure plotted along a straight line passing through
the Dirac point and parallel to $X$--$\Gamma$--$X$ (hence
labeled ($X'$--$\Gamma'$--$X'$).  The crossing is clearly linear
sufficiently close to the Dirac point at the center, with
quadratic and higher variations further from $\Gamma'$.  By contrast,
for the smaller P4$_{2}$/nmc unit cell proposed
previously,\cite{Z.Phys.Chem.B28.427.1935} we find that the
corresponding curve has almost no linear component, with
quadratic and higher behaviors dominating even very close to
the Dirac point.

Figures~\ref{fig:dft_mag}(c) and (d) show the corresponding
results in the presence of the effective Zeeman field of $h=100$\,T
along the $z$ axis.  The Kramers-degenerate bands are now all
split by the Zeeman field.
Along the $\Gamma$--$Z$ direction shown in \fref{dft_mag}(c), the
crossing bands belong to different irreducible representations
of the $C_4^z$ rotation operator, so that there is no avoided
crossing.  Instead, each of the four crossings generates a
Weyl node. The structure in the vicinity of these crossings is
shown in the inset of \fref{dft_mag}(c), where the four crossing
bands are shown in red, green, orange, and blue in order of
increasing energy.  The four Weyl nodes generated from
the crossings are labeled as $w_j$, with
$w_1$ connecting the
bottom two bands, $w_2$ and $w_3$ connecting the middle bands, and
$w_4$ connecting the top two bands.
Figure~\ref{fig:dft_mag}(d) shows the dispersion on the same
$X'$--$\Gamma'$--$X'$ line as in panel (b); this line does not
pass exactly through any of the Weyl points, so the states are
all nondegenerate at $\Gamma'$.  The dispersion looks like two
copies of panel (b), slightly shifted in energy by the Zeeman
perturbation, and with each showing an avoided crossings at
$\Gamma'$ because of the influence of SOC.
The details of the structure in the vicinity of the avoided crossings
in \fref{dft_mag}(d) will become clearer in the context of
the effective \kdotp model that we introduce next.

\begin{figure}
\includegraphics[scale=0.36]{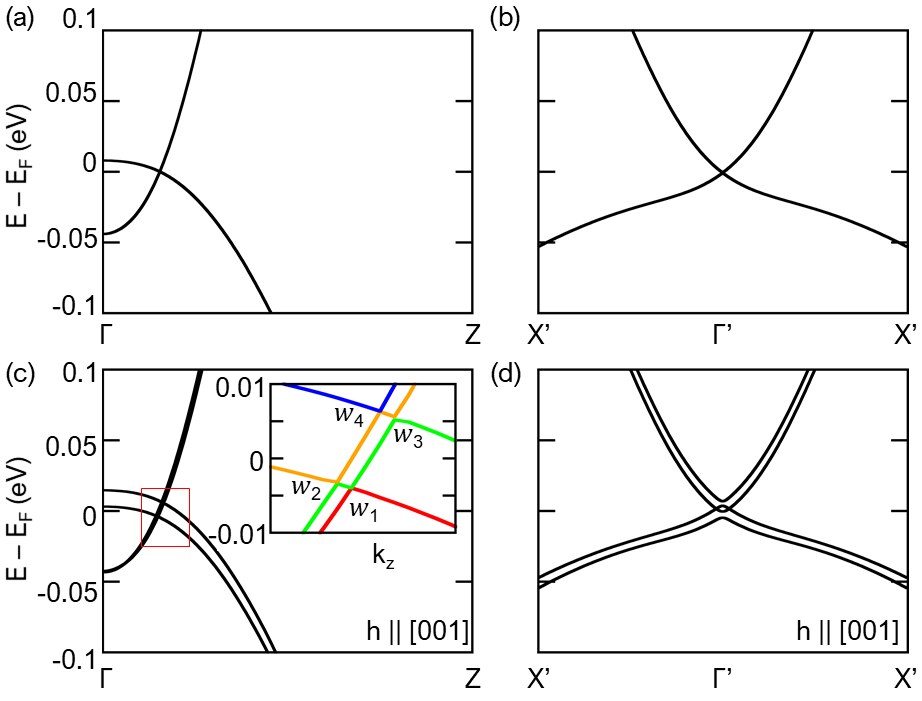}
\caption{
(a) Nonmagnetic PBE+SOC band structure of crossing Dirac bands along the
$\Gamma-Z$ direction.
(b) Same but along $X^{\prime}-\Gamma^{\prime}-X^{\prime}$
at $k_z=k_{\rm D}$, i.e., passing through the Dirac point.
(c-d) Same as (a-b), but under a Zeeman field of $h=100$\,T along
the [001] direction.
Inset of (c) shows four Weyl nodes labeled as $w_1$ through $w_4$.}
\figlab{dft_mag}
\end{figure}

\subsection{Effective \kdotp Hamiltonian}
%
To understand the behavior of this system and
compute its anomalous Hall conductivity for Fermi level positions
in the vicinity of the Weyl nodes, it is useful to introduce
a minimal effective \kdotp model for the bands in this
region.  Following the work of Wang and
collaborators,\cite{PhysRevB.85.195320,PRB.88.125427.2013}
%
%
our effective Hamiltonian is written in the basis of spin-orbit
coupled states $\ket{S_{1/2},J_{z}=1/2}$, $\ket{P_{3/2},J_{z}=3/2}$,
$\ket{S_{1/2},J_{z}=-1/2}$, and
$\ket{P_{3/2},J_{z}=-3/2}$, where $s$ and $p$ states reside on Cd
and As atoms respectively.  After defining $\k$ relative to the
Dirac point at $(0,0,k_{\rm D})$, expanding in powers of $\k$, and keeping
the leading terms allowed by symmetry, the Hamiltonian takes
the form
%
%
\begin{align}
&H({\bf k}) =\notag\\
&\begin{pmatrix}
v_{s}k_{z}+\beta_{s}h & Ak_{+} & 0 & Gk_{-}^{2} \\
Ak_{-} & -v_{p}k_{z}+\beta_{p}h & Gk_{-}^{2} & 0 \\
0  & Gk_{+}^{2}  & v_{s}k_{z}-\beta_{s}h & -Ak_{-}  \\
Gk_{+}^{2} & 0 & -Ak_{+} & -v_{p}k_{z}-\beta_{p}h
\end{pmatrix} .
\eqlab{ham}
\end{align}
%
%
The terms involving $v_{s}$, $v_{p}$, and $A$ describe a slightly
tilted Dirac cone with perfectly linear dispersion, where $v_s$ and
$v_p$ are the magnitudes of the Fermi velocities for the $s$ and
$p$ bands respectively, and $A$ determines the Fermi velocity in
the $k_x$ and $k_y$ directions, where $k_{\pm}=k_{x} \pm i k_{y}$.
Parameters $\beta_s$ and
$\beta_p$ represent the effect of the Zeeman exchange field $h$ on the two
sets of states.  Quadratic terms involving $k_z^2$ and $k_x^2+k_y^2$
have been omitted on the grounds that they will not be important
when working in a small region of $(\k,E)$ space close to the
Dirac point, and because they do not induce any Berry curvature.

By contrast, we have included the quadratic terms involving
$Gk_+^2$ and $Gk_-^2$.  These represent spin-orbit coupling, and
have important qualitative and quantitative effects on the nature
of the Fermi surfaces and the anomalous Hall response. Without
these terms, the upper-left and lower-right 2$\times$2 blocks
of $H(\k)$ (the ``spin up'' and ``spin down'' sectors) would be
completely uncoupled, leading to the existence of nodal loops
where spin-up and spin-down Fermi surfaces intersect.
Wieder et al.\cite{wieder_strong_2020} have pointed out (see their
Supp.~Eq.~237) that the four-fold rotational symmetry actually
allows the $Gk_+^2$ and $Gk_-^2$ terms to be generalized
to $Gk_-^2+G'k_+^2$ and $Gk_+^2+G'k_-^2$ respectively, with
$G'\ll G$. The inequality arises because $G'$, unlike $G$, would
vanish in the presence of continuous rotational symmetry, and is
only induced by an additional weak crystal field perturbation.
We therefore neglect the $G'$ term in this work.

The parameters $A$, $G$, $v_{s}$, $v_{p}$, $\beta_{s}$ and
$\beta_{p}$ are computed from the PBE+SOC band structure calculations to
make our microscopic model Hamiltonian close to the realistic picture.
The parameter $A$ is taken from the slope of the
bands at the Dirac crossing plotted in the ($k_x,k_y$) plane at
$k_z-k_{\rm D}$ in the absence of an external field,
as shown in the Fig.~\ref{fig:dft_mag}(b).
The parameters $v_{s}$ and $v_{p}$ are given by the slopes of the Cd $-s$
and As $-p$ bands at their crossing point, as in Fig.~\ref{fig:dft_mag}(a).
The parameters $\beta_{s}$ and $\beta_{p}$
describe the linear dependence of the exchange splittings of the
Cd~$s$ and As~$p$ bands on the strength of the Zeeman field $h$,
as determined by the band splittings very close to the Weyl nodes.

The prefactor $G$ of the off-diagonal
$k_{+}^{2}$ and $k_{-}^{2}$ terms is obtained from a close inspection
of the bands near $E=E_{\rm F}$ in Fig.~\ref{fig:dft_mag}(d), where
a tiny gap (not visible in the figure) arises at each crossing between
the second and third bands.  In the absence of $G$, the ``spin-up''
and ``spin-down'' sectors would become completely decoupled,
\eq{ham} would become block diagonal, and these avoided crossings
would disappear. Thus, a nonzero $G$ is required for a qualitatively
correct description. However, the determination of $G$ is rather
sensitive to details of the first-principles band structure, so
we later allow it to vary in order to study how these
avoided crossings affect the anomalous Hall conductivity.

The parameter values obtained as described above from the
first-principles calculations are
$A$ = 0.99\,eV-\AA,
$v_{s}$ = 2.68\,eV-\AA,
$v_{p}$ = 0.56\,eV-\AA,
$\beta_{s}$ = 0.054\,meV/T,
$\beta_{p}$ = 0.115\,meV/T,
and $G$ = 10\,eV-\AA$^2$.
The last three were obtained from calculations at $h$\,=\,100\,T,
but their values are not sensitive to variations of $h$ in this range.

\subsection{Chirality and Chern numbers of Fermi pockets}
%
The \kdotp model described above allows for a convenient
description of the locations of the Weyl points, the shapes of
the Fermi surfaces, and the behavior of the anomalous Hall
conductivity in the vicinity of the Dirac crossing.

As discussed earlier and illustrated in the inset of \fref{dft_mag}(c),
the single Dirac crossing in the absence of Zeeman field produces
a set of four Weyl points on the $k_z$ axis in the presence of
the field. Within our \kdotp model, the bands are exactly linear
along the $k_z$ axis as illustrated in \fref{fermi_pocket}(a),
and the locations $k_{z,j}$ of the Weyl points $w_1\,...\,w_4$ are
\begin{align}
k_{z,1}&=-(\beta_{p}-\beta_{s})h/(v_{s}+v_{p})\,, \nonumber\\
k_{z,2}&=-(\beta_{s}+\beta_{p})h/(v_{s}+v_{p})\,, \nonumber\\
k_{z,3}&=(\beta_{s}+\beta_{p})h/(v_{s}+v_{p})\,, \nonumber\\
k_{z,4}&=(\beta_{p}-\beta_{s})h/(v_{s}+v_{p})\,.
\end{align}
The corresponding chiralities are $\chi_1=-1$, $\chi_2=-2$,
$\chi_3=+2$, and $\chi_4=+1$, as obtained analytically
\cite{PhysRevB.96.045102,PhysRevLett.108.266802,PRL.107.186806.2011}
from the model Hamiltonian.
Our sign convention is such that a Weyl node of positive chirality
is a source and sink of Berry curvature in the conduction and
valence bands respectively.\footnote{This is the opposite
  sign convention from that adopted in Ref.~[\onlinecite{PhysRev.92.085138}]
  and some other works.}
As a reminder, $w_1$ connects the bottom two bands, $w_2$ and $w_3$
connect the middle bands, and $w_4$ connects the top two bands of
the four-band group.  With our model parameters,
Weyl points $w_1$ and $w_4$ occur at
$k_z=\mp1.9\!\times\!10^{-3}\,$\AA$^{-1}$ and $E=\mp10.5$\,meV,
and $w_2$ and $w_3$ occur at
$k_z=\mp5.2\!\times\!10^{-3}\,$\AA$^{-1}$ and $E=\mp8.6$\,meV.

%
\begin{figure}
\centering\includegraphics[height=3.7cm]{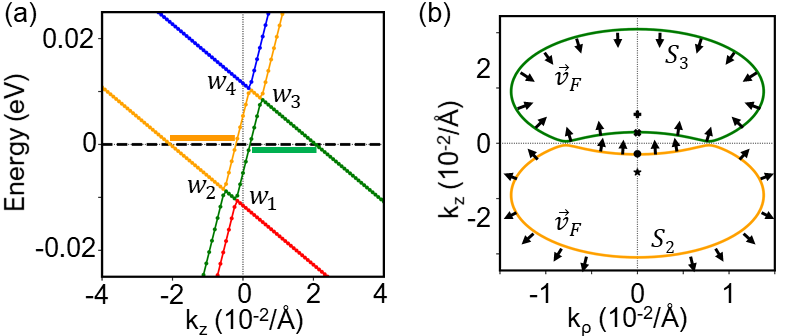}
\caption{
(a) Dispersion of the four Dirac-derived bands along the
$k_z$ axis within the \kdotp model.
Bands are colored red, green, orange, and blue in order of
increasing energy; Weyl points $w_1$, $w_2$, $w_3$, and $w_4$
are marked.
(b) Electron pocket in third band (orange) and hole pocket in
second band (green) for $E_F=0$, as indicated by
horizontal orange and green lines in (a).  Weyl point positions
are shown by dots on the $k_z$ axis.}
\figlab{fermi_pocket}
\end{figure}

Next, we vary the Fermi level over the energy range of the Weyl points
and study the evolution of the Fermi surfaces.  For this purpose it is
convenient to transform to cylindrical $(k_{\rho},k_{\phi},k_{z})$
coordinates, with $k_{\rho}^2=k_{x}^{2}+k_{y}^{2}$ and
$k_{\phi}=\tan^{-1}(k_{y}/k_{x})$.  The symmetry of \eq{ham} is such
that $E({\bf k})=E(k_\rho,k_z)$ independent of $k_\phi$, so all
Fermi surfaces have cylindrical symmetry within this model, and it
is convenient to plot Fermi surfaces in $(k_\rho,k_z)$ space.
For example, \fref{fermi_pocket}(b) shows the Fermi surfaces for the
case $E_{\rm F}=0$, the nominal charge neutrality point.  (The
full 3D Fermi surface would be obtained by rotating this figure
about the $k_x$ axis.) Referring back to \fref{fermi_pocket}(a),
it is clear that
the orange Fermi surface is an electron pocket surrounding occupied
Weyl point $w_2$, while the green one is a hole pocket surrounding
unoccupied Weyl node $w_3$. The small gap separating the orange and
green Fermi surfaces in \fref{fermi_pocket}(b) is a consequence of
the nonzero $G$ parameter in our model. Corresponding figures for
a range of Fermi energies spanning over the range of Weyl points are
shown in \fref{fermi_chern}.

%
\begin{figure}
\centering\includegraphics[height=10.8cm]{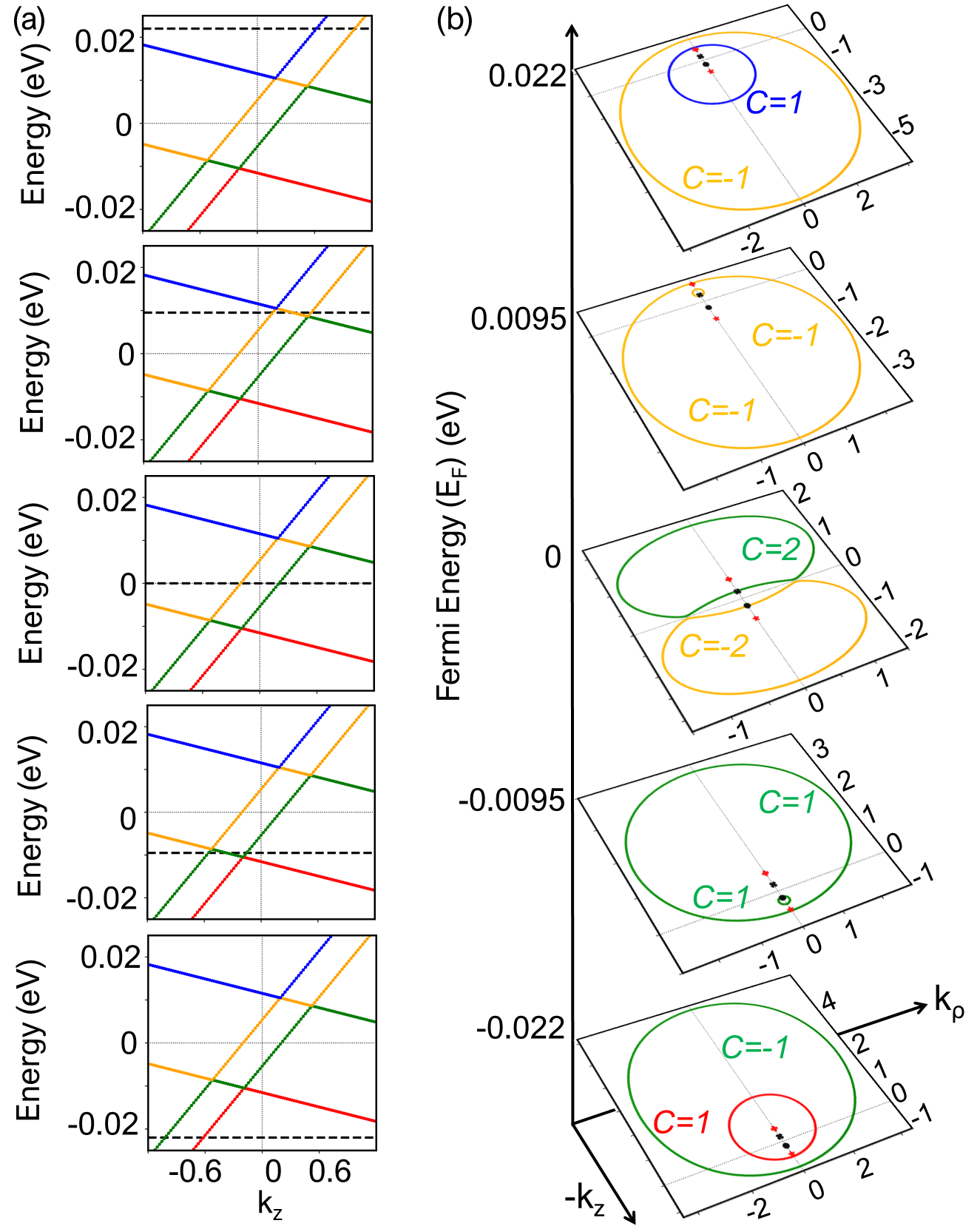}
\caption{(a) Band structure showing several Fermi-level positions (dashed
lines) as $E_F$ is tuned across the energy range of the Weyl nodes.
(b) Contours of Fermi surfaces projected on the
$(k_{z},k_{\rho})$ plane corresponding to each Fermi-level
position, for $G=10$\,eV\AA$^2$, with $k_{z}$ and $k_{\rho}$
in units of $10^{-2}$\AA$^{-1}$.  Chern numbers of electron and
hole pockets are indicated.}
\figlab{fermi_chern}
\end{figure}

The Chern number $C_n$ of a Fermi pocket band in band $n$ is obtained by
calculating the Berry flux passing through its Fermi surface
\cite{PRB.76.195109.2007,PRL.93.206602.2004} according to
\begin{equation}
C_{n}=\frac{1}{2\pi} \oint_{S_F} d^{2}k \, \Omega_{n}({\bf k}) \,,
\eqlab{flux-calc}
\end{equation}
where
\begin{equation}
  \Omega_{n}=\bm{\Omega_{n}}\cdot\hat{\bf v}_{F}
\eqlab{bc-normal}
\end{equation}
is the surface-normal component of the Berry curvature.
Note the sign convention:
$\hat{\bf v}_{F}$ is the Fermi velocity unit vector, which
is outward-directed for electron pockets and inward for hole pockets
as illustrated in Fig.~\ref{fig:fermi_pocket}(a).
The Berry curvature components in tensor notation
($\Omega_{\alpha\beta}=\epsilon_{\alpha\beta\gamma}\Omega_\gamma$)
are computed using the Kubo formula
\begin{equation}
\Omega_{\alpha\beta,n} = 2{\rm Im}\sum_{m\neq n}
  \frac{\bra{n}{v_\alpha}\ket{m}\bra{m}v_\beta\ket{n}}{(E_{n}-E_{m})^{2}} 
\eqlab{kubo}
\end{equation}
where $v_\alpha=\partial H/\partial k_\alpha$ are the velocity operators.
In practice we carry out the calculation in cylindrical coordinates,
computing $\Omega_z$ from matrix elements of $v_\rho$ and $v_\phi$
and $\Omega_\rho$ from matrix elements of $v_\phi$ and $v_z$.
To evaluate \eq{bc-normal} we need to combine these as
$\Omega = (\Omega_\rho v_{F,\rho}+\Omega_z v_{F,z})/|{\bf v}_F|$
Then the Fermi surface integral is carried out by discretizing
each $(k_\rho,k_z)$ path describing a Fermi surface, such as the
green curve in \fref{fermi_pocket}(a).  We then
compute the needed ingredients at each $(k_\rho,k_z)$ point along
the path and sum, taking account of phase space details such
as the $2\pi$ coming from the $k_\phi$ integral.
We have checked that the results are always very close to integer values
as long as the discrete sampling is sufficiently dense.

The resulting Fermi surface Chern numbers are shown for a range of
Fermi level positions in Fig.~\ref{fig:fermi_chern}.  The Fermi
level positions are indicated by black dashed lines in
\fref{fermi_chern}(a) and the corresponding Fermi surfaces projected
onto the $k_{\rho}-k_{z}$ plane are shown in
Fig.~\ref{fig:fermi_chern}(b).

Starting from the top of Fig.~\ref{fig:fermi_chern}, where the
Fermi level crosses only the blue and orange bands, the Chern numbers
are $+1$ and $-1$ for these bands respectively. This is consistent with
Gos\'albez-Mart\'{\i}nez et al.,\cite{PhysRev.92.085138}
where the Chern number $C_n$ of a Fermi surface in band $n$
was shown to be equal to the sum of chiralities
of the enclosed Weyl points connecting to band $n-1$ minus the
corresponding sum for touchings with band $n+1$.
For the blue band there is only a single contribution of the first
type, coming from $w_4$ with $\chi_4=+1$, giving $C_4=+1$. For the
orange band $w_2$ and $w_3$ contribute positively and $w_4$
contributes negatively, giving $C_3= (-2) + (+2) - (+1) = -1$.

The remaining panels of \fref{fermi_chern} show the evolution
of the Fermi surfaces as the Fermi energy is swept through the
region of the Weyl points.  We find topologically nontrivial
Fermi pockets in all cases, suggesting that Cd$_{3}$As$_{2}$
may be a promising material for realizing
unconventional superconductivity based on topological
Fermi surfaces in a Weyl semimetal.\cite{PhysRevLett.120.067003,sun-lee-li}
When the Fermi level is at the nominal charge neutrality level
of $E_F=0$, we find electron (orange) and hole (green) pockets with
Chern numbers $\mp 2$ respectively.
The effect of the finite $G$=10\,eV\AA$^2$
makes a small separation between these pockets,
as shown in the Fig.~\ref{fig:fermi_chern}(b),
with a separation that grows larger as $G$ is increased.
In the next section we investigate the effect of varying this
parameter on the Berry curvature and anomalous Hall conductivity
of the system.

\subsection{Berry curvature}
%
As we shall see, the parameter $G$ in the \kdotp Hamiltonian plays
an important role in producing
Berry curvature and influencing the anomalous Hall conductivity.
However, even when $G=0$, Berry curvature is present.  In this case,
the separation between the electron and hole pockets in
Fig.~\ref{fig:fermi_chern}(b) for $E_F$ near zero vanishes.
In this case the
spin-up and spin-down sectors of \eq{ham} do not mix, and their
ellipsoidal Fermi surfaces intersect without any avoided crossing.
Nevertheless, there is still a nonzero Berry curvature, since
Weyl points $w_1$ and $w_4$ are still present and serve as equal
and opposite sources of Berry flux in the two spin sectors.
The fact that these are offset from one
another along the $k_z$ axis allows for a nonzero net anomalous
Hall conductivity.

%
\begin{figure}
\centering\includegraphics[height=7cm]{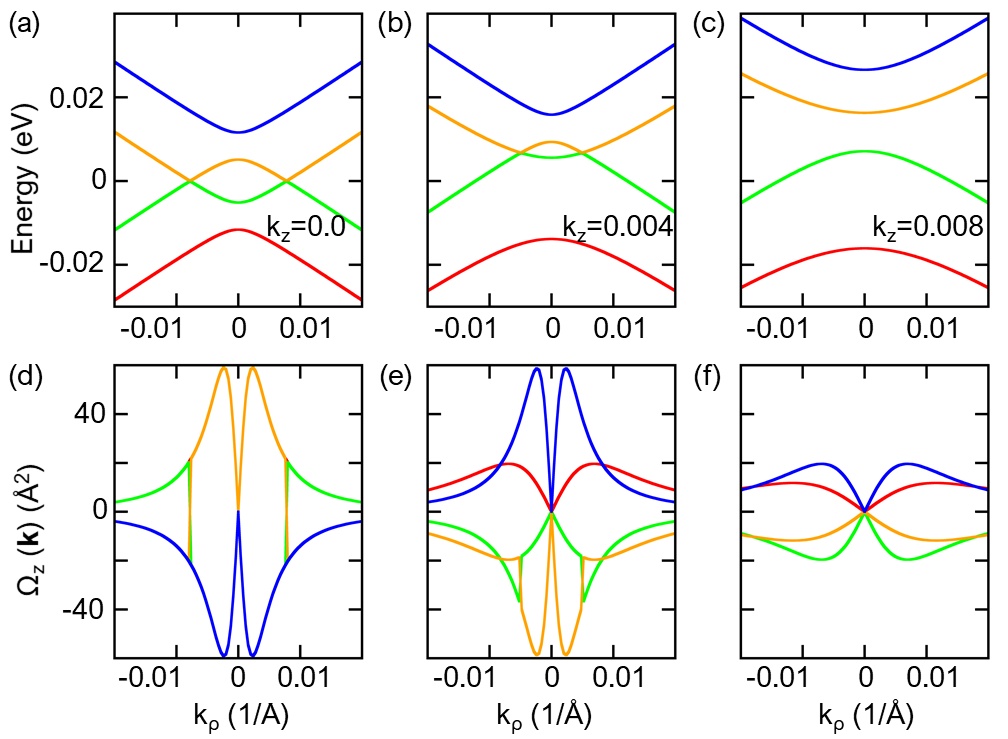}
\caption{
(a-c) Band structure plotted versus $k_\rho$ for three
different values of $k_{z}$ when $G$ is zero.
(d-f) Berry curvature of the corresponding bands, color-coded accordingly.
}
\figlab{gzerobc}
\end{figure}

\Frefs{gzerobc}(a-c) shows the band structure plotted versus
$k_\rho$ for three values of $k_z$.  At the critical
value $k_z=k_{z,3}=0.0052$\,\AA\ (not shown), the second and third
band have a quadratic touching; this will become the location
of Weyl point $w_3$, but with $G=0$ it is still part of a surface
of degeneracy between the second and third bands.
The corresponding Berry curvature of the bands is plotted in
\frefs{gzerobc}(d-f).

%
\begin{figure}
\centering\includegraphics[height=7cm]{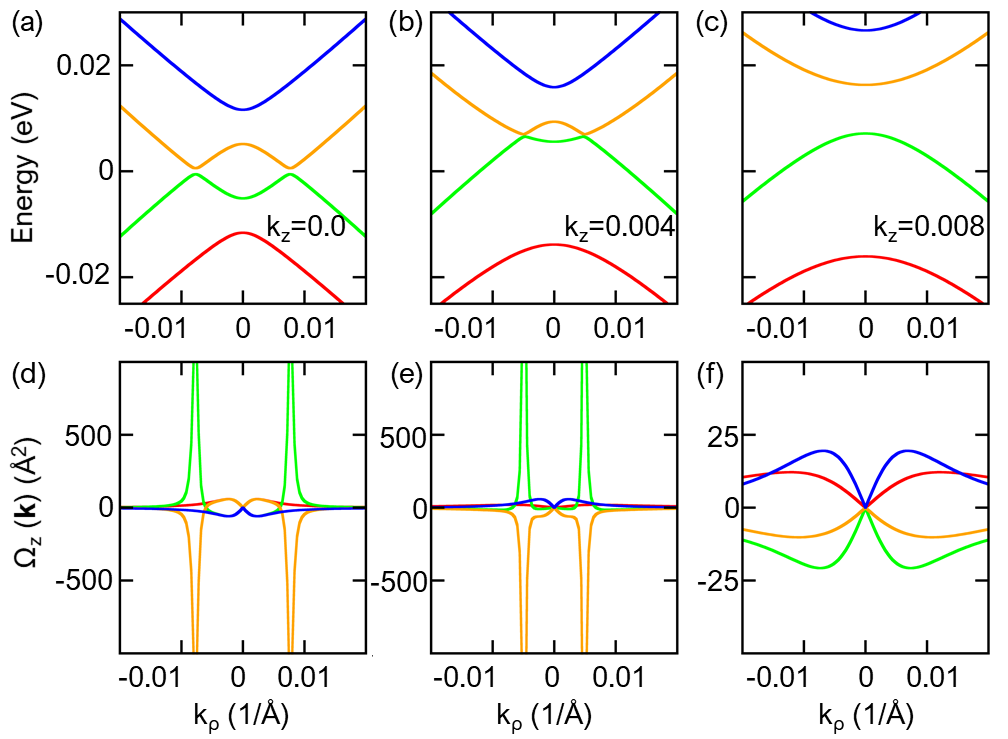}
\caption{
(a-c) Band structure plotted versus $k_\rho$ for three
different values of $k_{z}$ when $G$ is $10$\,eV\AA$^2$.
(d-f) Berry curvature of the corresponding bands, color-coded accordingly.
}
\figlab{gbc}
\end{figure}

Turning on a finite $G$ causes a mixing of the spin-up and spin-down
sectors, so that crossings between bands associated with these
sectors are gapped almost everywhere.  The exceptions are the
locations of the quadratic Weyl points $w_2$ and $w_3$, which
survive the arrival of the finite $G$.  Because $G$ multiplies
$k_+^2$ and $k_-^2$ terms in \eq{ham}, it is responsible for
the higher-order ($\chi=\pm2$) nature of these Weyl points.
The band structures are shown in \frefs{gbc}(a-c) for a 
$G$ value of $10$\,eV\AA$^2$. There are now small avoided crossings
between the second and third bands in \fref{gbc}(a-b).
\Frefs{gbc}(d-f) show the corresponding
Berry curvature on these bands, showing very large
peaks near the avoided crossings, with the potential to make
large contributions to the anomalous Hall conductivity.

\subsection{Anomalous Hall conductivity}
%
Once the time-reversal symmetry is broken by the presence of the
Zeeman field, a nonzero Hall conductivity
$\sigma_{yx}^{\rm AHC}$ is expected on symmetry grounds.
We refer to this as ``anomalous'' Hall conductivity since
we have in mind that the magnetic order has its origin in
magnetic impurities or proximity effects. We note in passing
that Hall conductivity measurements have been reported for
Cd$_{3}$As$_{2}$ nanoplates.\cite{Zhang2019,Zhang2017}
Here we report the the behavior of the intrinsic anomalous Hall conductivity
as a function of Fermi level position in the context of our
\kdotp model of bulk Cd$_{3}$As$_{2}$.

To calculate the intrinsic anomalous Hall conductivity, we have 
adopted the Fermi-surface integral approach of
Refs.~[\onlinecite{PRL.93.206602.2004,PRB.76.195109.2007}].
In this formulation
\begin{equation}
\sigma_{\alpha\beta}=\frac{-e^{2}}{\hbar}\frac{1}{2\pi^{2}}\sum_{n\gamma}
\epsilon_{\alpha\beta\gamma} K_{n\gamma} 
\,,
\end{equation}
where 
\begin{equation}
K_{n\gamma}=\frac{1}{2\pi} \oint_{S_F} d^{2}k \, \Omega_{n}\, k_\gamma
\eqlab{K-calc}
\end{equation}
measures the surface-normal Berry flux passing through the Fermi surface,
as in \eq{flux-calc}, but now weighted by the wavevector component $k_\gamma$.
As the $z$ component of the intrinsic anomalous Hall conductivity (i.e.,
$\sigma_{yx}^{\rm AHC}$) is the
only one allowed by symmetry, we compute $K_{nz}$ for each band $n$ and
for each point on the Fermi surface and integrate, as described
previously below \eq{kubo} for the earlier Chern-number calculation.

%
\begin{figure}
\centering\includegraphics[height=6.2cm]{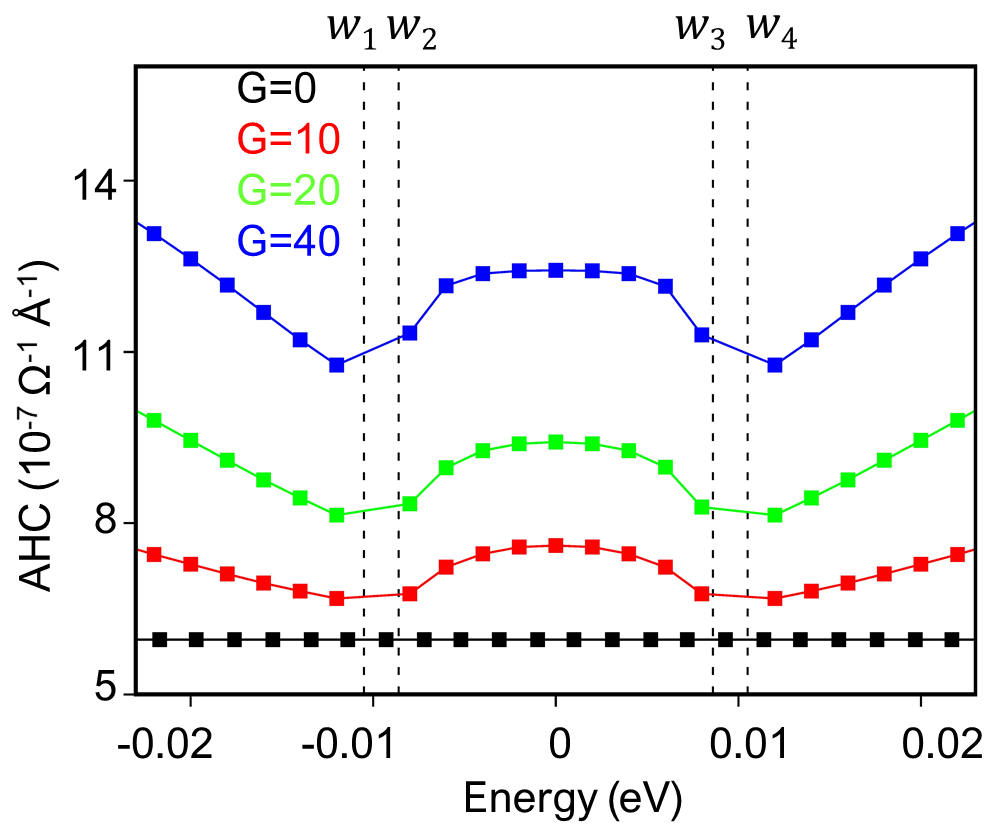}
\caption{
Anomalous Hall conductivity $\sigma_{yx}^{\rm AHC}$ plotted versus Fermi
level position for several values of the $G$ parameter.
Energies of the Weyl nodes are indicated by the
vertical dashed lines.}
\figlab{ahc}
\end{figure}

The resulting anomalous Hall conductivity is plotted as a function of
Fermi-level position for several values of the $G$ parameter
in Fig.~\ref{fig:ahc}, again for our reference Zeeman field of $h$=100\,T.
The analysis above applies to the split Dirac cone
located on the positive $k_z$ axis, but there is an equal
contribution coming from the inversion image on the negative
$k_z$ axis (inversion does not reverse the sign of the anomalous
Hall conductivity), giving an additional factor of two that has
been included in the results presented in Fig.~\ref{fig:ahc}.

When $G=0$, we find a result that is nonzero but constant as
a function of Fermi level position.  Further investigations shows that
the spin-up and spin-down contributions are both perfectly linear
in $E_F$, as expected for tilted Weyl cones, but the slopes are
equal and opposite so that the sum is constant.  However, these
contributions do not simply cancel; the constant residual can be
understood as coming from the off-centering along $k_z$ of the
spin-up and spin-down Weyl cones.

Turning now to the results for nonzero $G$, we find that an
additional contribution to the anomalous Hall conductivity grows
in with increasing $G$.  This contribution follows an interesting
pattern in which there is an inverted dome in the energy region between
the upper and lower pairs of Weyl nodes, a relatively smaller
contribution near those nodes, and then a further growth that is
roughly linear in $|E_F|$ in the energy range outside the nodes.
We have varied the effective Zeeman field value from
$100$\,T to $400$\,T and obtained a qualitatively similar
behavior.
This distinctive behavior of the anomalous Hall conductivity could
serve as a fingerprint of a material with a Zeeman-split
Dirac cone if it can be observed experimentally, as by transport
measurements of a gated thin film.

\section{Summary and conclusions}

Motivated to understand the effect of magnetic order arising from
magnetic doping or proximity in the Dirac semimetal Cd$_{3}$As$_{2}$,
we have presented a detailed investigation of the splitting of the
Dirac node into Weyl nodes in the presence of
a time-reversal breaking Zeeman field oriented along the symmetry axis.
We emphasize that the Dirac node does not simply split into
a pair of Weyl nodes with chirality $\pm1$ as is commonly expected.
Instead, we find two nodes of chirality $\pm2$ connecting the nominal
valence and conduction bands, in addition to nodes of chirality
$\pm1$ connecting lower and higher pairs of bands.
Starting from first-principles density-functional calculations and
fitting a \kdotp model that is well suited to explore the low-energy
physics, we analyze the evolution of the Fermi surfaces, their Chern
numbers, and their contributions to the anomalous Hall conductivity,
as a function of Fermi level position using the \kdotp model.
The behavior of the anomalous Hall conductivity shows a distinctive
pattern that may be a suitable target for experimental confirmation.
The presence of multiple, topologically nontrivial electron and hole
pockets for Fermi energies near charge neutrality suggests possible
avenues for the realization of novel forms of superconducting
pairing. Finally, we anticipate that the methods presented here can
find use more generally in unraveling the intriguing physics of
time-reversal symmetry breaking in Dirac semimetals.

\acknowledgments{This work was supported as part of the Institute
for Quantum Matter, an Energy Frontier Research Center funded by
the U.S.\ Department of Energy, Office of Science, Basic Energy
Sciences under Award No.~DE-SC0019331.  We thank Yi Li for
useful discussions.}

\bibliography{pap}

\end{document}